\documentclass{new_tlp}

\title[Towards Assertion-based Debugging of Higher-Order (C)LP Programs]
  {Towards Assertion-based Debugging of \\ Higher-Order (C)LP Programs
  \thanks{Research supported in part by projects EU
    FP7 318337 \emph{ENTRA}, Spanish MINECO TIN2012-39391
    \emph{StrongSoft} and TIN2008-05624 \emph{DOVES}, and Comunidad de
    Madrid TIC/1465 \emph{PROMETIDOS-CM}.}
}

\author[Nataliia Stulova, Jos\'{e} F. Morales,
  and Manuel V. Hermenegildo]{
       NATALIIA STULOVA$^{1}$ 
    ~~ JOS\'{E} F. MORALES$^{1}$ 
    ~~ MANUEL V. HERMENEGILDO$^{1,2}$ 
\ \\
\ \\
   $^1$IMDEA Software Institute \\
   \email{\{nataliia.stulova, josef.morales, manuel.hermenegildo\}@imdea.org} \\
   $^2$School of Computer Science, Technical University of Madrid (UPM) \\
   \email{manuel.hermenegildo@upm.es}\\ 
}

\submitted{February 4, 2014}
\revised{March 18, 2014}
\accepted{March 18, 2014}

\begin{document}
\maketitle

\centerline{\textbf{Extended Abstract}}

\ \\
Higher-order programming adds flexibility to the software development
process. Within the (Constraint) Logic Programming ((C)LP) paradigm,
Prolog has included higher-order constructs since the early days, and
there have been many other proposals for combining
the first-order kernel of (C)LP with different higher-order
constructs. 
Many of these proposals are currently in use in different
(C)LP systems and have been found very useful in programming practice,
inheriting the well-known benefits of code reuse (templates),
elegance, clarity, and modularization.

A number of extensions have also been proposed for (C)LP in order to
enhance the process of error detection and program validation. In
addition to the use of classical strong typing, 
a number of other
approaches have been proposed which are based on the dynamic and/or
static checking of user-provided, optional \emph{assertions}.
Of these, the model of~\cite{ciaopp-sas03-journal-scp} has
perhaps had
the most impact in practice and different aspects of this model have
been incorporated in a number of widely-used (C)LP systems, such as
Ciao, SWI, and XSB.
A similar evolution is represented by the soft/gradual typing-based
approaches in functional programming and the contracts-based
extensions in object-oriented programming.



These two aspects, assertions and higher-order, are not independent.
When higher-order constructs are introduced in the language it becomes
necessary to describe properties of arguments of predicates/procedures
that are themselves also predicates/procedures.  While the combination
of contracts and higher-order has received some attention in
functional programming,
within (C)LP the combination of higher-order with the previously
mentioned assertion-based approaches has received comparatively little
attention to date. Current Prolog systems
simply use
basic atomic types (i.e., stating simply that the argument is a
\texttt{pred}, \texttt{callable},
etc.) to describe
predicate-bearing variables.
%
%
Other approaches are oriented instead 
to meta programming,
describing meta-types 
but 
there is no notion of directionality (modes), and only a single
pattern is allowed per predicate.

Our work~\cite{asrHO-tr} contributes towards filling this gap between
higher-order (C)LP programs and assertion-based extensions for error
detection and program validation.  
Our starting point is the Ciao assertion model, since, as mentioned
before, it has been adopted at least in part in a number of the most
popular (C)LP systems.

We have proposed an extension of the traditional notion of programs
and derivations in order to deal with higher-order calls and we have
adapted the notions of first-order conditional literals, assertions,
program correctness, and run-time checking to this type of
derivations.  This has allowed us to revisit the traditional model in
this new, higher-order context, while introducing a different
formalization than the original one, which is more compact and gathers
all assertion violations as opposed to just the first one, among other
differences.
%
%
We have defined an extension of the
%
%
properties used in assertions and of the assertions themselves to
higher-order, and provided corresponding semantics and results.

We have defined a new class of properties, ``predicate properties''
(\emph{predprops} in short), and proposed a syntax and semantics for
them. These new properties can be used in assertions for higher-order
predicates to describe the properties of the higher-order arguments.
%
%
%
We have also proposed several operational semantics for performing
run-time checking of programs including predprops and provided
correctness results for them.

Our predprop properties specify conditions for predicates that are
independent of the usage context. This corresponds in functional
programming to the notion of \emph{tight} contract
satisfaction, and it contrasts
with alternative approaches such as \emph{loose} contract
satisfaction.
In the latter, contracts are attached to higher-order arguments by
implicit function wrappers. The scope of checking is local to the
function evaluation. Although this is a reasonable and pragmatic
solution, we believe that our approach is more general and more
amenable to combination with static verification techniques.
%
For example, avoiding wrappers allows us to remove checks (e.g., by
static analysis) without altering the program semantics.
%
%
Moreover, our approach can easily support \emph{loose} contract
satisfaction, since it is straightforward in our framework to
optionally include wrappers as special predprops.

We have included the proposed predprop extensions in an experimental
branch of the Ciao assertion language implementation. This has the
immediate advantage, in addition to the enhanced checking, that it
allows us to document higher-order programs in a much more accurate
way. 
We have also implemented several prototypes for operational semantics
with dynamic predprop checking,
which are being integrated into the already existing assertion
checking mechanisms for first-order assertions. Finally, we are
developing analyses for static verification of assertions containing
predprops.

\bibliographystyle{acmtrans}

\end{document}